\documentclass[ag]{copernicus}

\usepackage{color}

\begin{document}

\title{Weibel, Firehose and Mirror mode Relations}

\author[1,2]{{R. A. Treumann}
}
\author[3]{{W. Baumjohann}}

\affil[1]{Department of Geophysics and Environmental Sciences, Munich University, Munich, Germany}
\affil[2]{International Space Science Institute, Bern, Switzerland}
\affil[3]{Space Research Institute, Austrian Academy of Sciences, Graz, Austria}

\runningtitle{Weibel, Firehose, Mirror modes}

\runningauthor{R. A. Treumann and W. Baumjohann}

\correspondence{R. A. Treumann\\ (rudolf.treumann@geophysik.uni-muenchen.de)}

\received{ }
\revised{ }
\accepted{ }
\published{ }


\firstpage{1}

\maketitle

\begin{abstract}
Excitation of Weibel magnetic fields in an initially non-magnetized though anisotropic plasma may trigger other low-frequency instabilities fed by pressure anisotropy. It is shown that under Weibel-like-stable conditions the Weibel-like thermal fluctuation magnetic field allows for restricted Firehose-mode growth. In addition, low frequency Whistlers can also propagate in the plasma under certain anisotropic conditions. When the Weibel-like mode becomes unstable, Firehose instability ceases but Mirror modes take over. This will cause bubble structures in the Weibel-like field in addition to filamentation. 

\end{abstract}

\introduction
{Historically, t}here are just three celebrated {fundamental} very low frequency (electro-)magnetic instabilities in hot anisotropic plasmas, the well known firehose mode \citep[][for a plasma physics textbook]{vedenov1961,treumann1997}, its complementary equivalent, the Mirror mode, and the Weibel instability \citep[][and others]{weibel1959,yoon1987}. {Recently, \citet{schlick2011a} and \citet{schlick2011b}, applying a substantially more rigorous relativistic approach based on the theory of analytical functions, identified a much larger number of different electromagnetic low-frequency modes, weakly and strongly damped/unstable ones, which add additional dispersion channels to a magnetized plasma. \citet{felten2013a,felten2013b,felten2013c} and \citet{felten2013} extended these calculations to relativistic Weibel-like (non-magnetic) conditions, again finding a large number of different dispersion channels which belong to damped or unstable modes including very low frequency and non-oscillating aperiodic modes. }

{Of the historical modes, t}he first is a general bulk plasma mode excited in an external magnetic field by a thermal anisotropy with larger magnetically parallel than perpendicular temperature, $T_\perp>T_\|$, resulting in Alfv\'en waves which radiate along the magnetic field. At the contrary, the latter {Weibel-like} mode acts in non-magnetized plasmas when, by some not further specified reason, the plasma exhibits a thermal (pressure) anisotropy with higher temperature (kinetic energy) in one than in the two other directions. In both the firehose and {Weibel-like} cases the cause of a higher \emph{equivalent} temperature in one direction (and thus a temperature anisotropy) can also be a fast (relative) streaming \citep[for physical mechanism of the Weibel-like streaming mode see][]{fried1959} of the plasma with kinetic energy exceeding the transverse thermal energy. This can be provided by  beam or counter-streaming beam configurations \citep[cf., e.g.,][]{achterberg2007} and has been made responsible for the generation of magnetic fields under various non-dynamo conditions occurring, for instance, in shock waves \citep[for a review see, e.g.,][]{treumann2009}, preferentially in relativistic shocks \citep[for relativistic shocks see the review by][]{treumann2011}. The physical differences in the two modes are large. While Weibel{-like modes} provide a non-dynamo mechanism to \emph{produce} quasi-stationary and thus non-propagating magnetic fields in an otherwise non-magnetized plasma, the firehose mode grows on an existing field and propagates along the field at Alfv\'en velocity $V_A$ thus transporting energy away from the region where it is excited, filling a large volume with magnetic fluctuations and contributing to turbulence and other effects. However, since both modes are generated by similar mechanisms though being different, having completely different properties, one expects that they will compete and possibly even act in tandem to generate magnetic fields and propagate them away from the source region. 

The remaining Mirror mode, at the contrary, is an about stationary feature of the plasma with wave vector $\mathbf{k}$ almost perpendicular to the ambient magnetic field. It generates plasma inhomogeneity at the lowest frequency of plasma turbulence. It is easily derived from anisotropic fluid theory, but its physical mechanism is attributed to trapping of particles in depleted magnetic field regions along the field. This mechanism is more complex and still not completely resolved \citep[for a recent more complete account of the electron mirror mode which concerns us here, see, e.g.,][and references therein]{pokhotelov2008,pokhotelov2010,pokhotelov2013}. 

In the present note we briefly examine this situation {at the example of the classical Weibel instability}, showing that there indeed exists such a competition which may become important in limiting the growth of the {Weibel-like modes}, allowing other fluctuations to propagate on the magnetic {Weibel-like} background field which may contribute to distribution of magnetic fields in a larger volume. {A similar analysis examining the newly identified electromagnetic modes \citep{schlick2011a,schlick2011b,felten2013,felten2013a,felten2013b,felten2013c}, though being highly desirable, lies outside this brief communication.}

\section{Thermal fluctuation effects}
{In order to demonstrate the role of thermal magnetic fluctuations as a pathway for other modes, we restrict to the conventional Weibel-like thermal-anisotropy mode only. The following analysis could be made more complete by applying the expressions for the many dispersion channels identified in the above given references on the non-magnetized plasma modes. Here we refrain from such an extension of the present communication, not at least for the reason of the wealth of newly found modes but also for the complexity of the more precise expressions given there \citep{felten2013,felten2013a,felten2013b,felten2013c} which would obscure our intendedly focussed discussion.}  

{The thermal anisotropy-excited Weibel-like mode grows under the condition} that the anisotropy is along direction $z$. Writing $T_z=T_\|, T_{x,y}=T_\perp$ the condition for growth is, {in its simplest form,} given by
\begin{equation}
A\equiv \frac{T_\|}{T_\perp}-1> 0
\end{equation}
In this case it generates a magnetic field $\mathbf{B}_W$ in the perpendicular direction, i.e. transverse to the direction of anisotropy. Under the opposite condition the Weibel mode is stable but still generates a zero-frequency perpendicular magnetic thermal fluctuation field \citep{yoon2007,treumann2012} of spectral energy density\footnote{Clearly, under the opposite condition the direction of anisotropy switches and the Weibel instability works in the orthogonal direction. It is thus \emph{universal, working at any thermal anisotropy in nonmagnetic plasma} and disappears only for $A\equiv 0$. Just for this reason it also makes sense to write the thermal Weibel level including a non-vanishing anisotropy, for the thermal level becomes itself anisotropic: it is thermal along the anisotropy and thermally excited transverse to it.}
\begin{equation}\label{eq-2}
\frac{\langle b^2(k)\rangle}{ b_0^2}=\frac{(A+1)^2k\lambda_e}{(A+2)\left[k^2\lambda_e^2-A-\mu\right]^2}
\end{equation}
written here for the electron-Weibel mode with skin depth $\lambda_e$ and mass ratio $\mu=m_e/m_i$. Its spectral density amplitude is given by 
\begin{equation}
\frac{b_0^2}{m_ec^2}=\frac{\mu_0}{m_e}\sqrt{\frac{\pi T_\perp}{m_ec^2}}
\end{equation}
The thermal fluctuation level has a distinct dependence on wave number $k$. For short wavelengths $k\gg\lambda_e^{-1}$ it decays like $\propto(k\lambda_e)^{-3}$ \citep[confirming the result of][]{yoon2007}. Due to the presence of the inert ion component it vanishes at $k\to 0$, maximizing below $k\lambda_e<1$, i.e. compared with the electron skin depth the fluctuation field is of longer scale, providing a weakly-oscillatory moderate-wavelength background magnetic field. One may note that the wavenumber is perpendicular to both, ${\hat z}$. i.e. the direction of thermal pressure anisotropy, and the Weibel-like magnetic fluctuation field. 

{The magnetic fluctuation field is quasi-stationary in the sense that in the final step of calculation the real part of the frequency $\omega_r\approx 0$ was put to zero, which is equivalent to the -- not entirely correct -- assumption of purely growing/damped $\omega(\mathbf{k})=i\gamma(\mathbf{k})$ wave modes with growth/damping rate $\gamma(\mathbf{k})$. In fact, $\omega_r\neq 0$ for Weibel-like modes. Hence, with $k_0^2\lambda_e^2=A$ the requirement that $|\gamma(\mathbf{k})|<\omega_r(\mathbf{k})$ implies a condition on the wave number 
\begin{equation}
1> k^2\lambda_e^2>A\left(1-\sqrt{\frac{\pi}{4A(A+1)^3}}\right)
\end{equation}
Under damped conditions $A=-|A|$, and from Eq. (\ref{eq-2}) one requires $1<|A|<2$ fixing the range of applicability of the following analysis. There is a sufficiently wide range of wavelengths longer than $\lambda_e$ in the damped case. In the unstable case, shorter wavelengths around $k\sim\lambda_e$ are favored. The more precise analysis \citep{felten2013a,felten2013b} is  not restricted to the condition $|\gamma|<\omega_r$ but includes aperiodically growing/damped waves.} 

The mean magnetic fluctuation field amplitude ${\bar b}=b_0\left[\int \mathrm{d}{k}\langle b^2(k)\rangle\right]^\frac{1}{2}$ that results from the above thermal spectral density when integrating over $k$-space, becomes
\begin{equation}
\frac{\bar b}{b_0}=\left[\frac{(A+1)^2(A+1+\mu)}{(A+2)(A+\mu)(1-A-\mu)}\right]^\frac{1}{2}
\end{equation}
It serves as background magnetization on which electromagnetic waves at low frequency, much less than the electron plasma frequency $\omega\ll\omega_e$, can propagate in a nonmagnetic plasma. Propagation would be inhibited otherwise, allowing only for evanescent modes in the spatial range of the electron skin depth $\lambda_e$. Actually, for the same reason the unstable Weibel-like mode can penetrate just over the electron skin depth only from its generation site into the collisionless plasma perpendicular to the direction of anisotropy. This restriction necessarily causes a pronounced magnetic filamentation of the Weibel-like unstable plasma. On the other hand, once Weibel-like thermal fluctuations provide a weak magnetic background field, spreading across the plasma becomes possible  for other electromagnetic modes.

\section{Secondary instability}

{Once the thermal fluctuation spectrum of the Weibel mode is established, the plasma behaves weakly magnetized with the Weibel-like thermal fluctuation background field being structured at about the electron skin depth in the direction perpendicular to the anisotropy.  This implies that the thermal magnetic background organizes into long magnetic filaments of typical transverse size of few electron skin depths with consequences for the propagation of any secondary electromagnetic modes. }

\subsection{Weibel-stable case}
Let us assume that the Weibel mode has been stable (see footnote), which implies that $A\lesssim0, 1<|A|<2$. Under this condition we have $T_\perp>T_\|$, with $T_\perp$ being directed along the Weibel magnetic field, i.e. playing the role of a \emph{magnetically parallel} anisotropy, while $T_\|$ is a perpendicular anisotropy. Under this condition the Firehose mode could grow only under the condition
\begin{equation}
\frac{\beta_\perp}{\beta_\|}-1>\frac{2}{\beta_\|}, \qquad \beta_{\|,\perp}=\frac{2\mu_0NT_{\|,\perp}}{{\bar b}^2}
\end{equation}
with ${\bar b}^2=\int dk\langle b^2(k)\rangle$ the thermal rms magnetic field (note that indexing is with respect to the direction of the initial thermal anisotropy where the directions with respect to the magnetic field are inverted!) existing in the otherwise non-magnetic plasma. The plasma-$\beta$s refer to it in the present case. The left-hand side is positive in the Weibel-stable case and, hence, the Firehose instability \emph{can} grow under the rewritten condition
\begin{equation}\label{eq-fire}
\beta_\perp -\beta_\|> 2, \qquad |A| < 2
\end{equation}
where the second condition is imposed by the thermal fluctuation level requirement. {The firehose mode can also propagate, because its propagation direction is along the thermal Weibel-like magnetic field which is along the direction of the Weibel-like magnetic channels and therefore allowed to propagate for a parallel mode like the Firehose-Alfv\'en wave.}

Checking for the electron Mirror mode (not including any more sophisticated effects) yields instability as long as
\begin{equation}\label{eq-mirr}
\beta_\|^2-\beta_\perp >\beta_\perp\beta_\| \qquad\mathrm{and}\qquad A=\beta_\|/\beta_\perp-1<0 
\end{equation}
This implies a contradiction thus \emph{excluding} the mirror mode. Weibel-stable plasmas are stable against  Mirror but allow for Firehose modes. These propagate on the thermal level of the magnetic field in the direction perpendicular to the Weibel-stable direction of anisotropy. This is interesting to know as it shows that in an otherwise non-magnetic plasma which in one direction (this time the $\perp$-direction) the plasma is Weibel \emph{unstable} the existence of thermal magnetic fluctuations in the Weibel-stable direction causes Firehose modes to radiate away from this region if only the above Firehose condition, based on the small thermal fluctuation level of the magnetic field, is satisfied. Since the thermal level is small this will, in praxis, always be the case if only $T_\perp>T_\|$.

In addition, if the plasma consists of a thermal background and a warm anisotropic component, the low-frequency Whistler (or Alfv\'en) mode can be destabilized as it only requires that for the anisotropic component $A<0$ and hence $\beta_\perp<\beta_\|$ which is given by the stability condition of the Weibel mode (note again that the indices $\|,\perp$ refer to the anisotropy frame, not to the magnetic frame!). The spectral density of the magnetic fluctuations in this range with $\beta_\|/\beta_\perp<1$ in the long wavelength range $k\lambda_e<1$ is estimated as
\begin{equation}
\frac{\langle b^2(k)\rangle}{ b_0^2}\approx \left(\frac{T_\|}{T_\perp}\right)^2k\lambda_e
\end{equation}
Using the Whistler dispersion relation for electrons in the low frequency range, i.e. $k^2\lambda_e^2\approx \omega/\left(\Omega_e-\omega\right)$, with $\Omega_e= e{\bar b}/m_e$ the electron cyclotron frequency in the quasi-stationary Weibel thermal fluctuation field, yields the relation
\begin{equation}
\frac{\langle b^2(k)\rangle}{ b_0^2}\approx \left(\frac{T_\|}{T_\perp}\right)^2\sqrt{\frac{\nu}{1-\nu}}, \qquad \nu=\frac{\omega}{\Omega_e}
\end{equation}
between the Whistler frequency range $\omega(k)$ and the magnetic spectral energy density. Since we know that the magnetic spectral density vanishes at $k=0$, and using the dispersion relation for $k\lambda_e=1$, we find that the frequency of Whistlers is in the range
\begin{equation}
0\leq\nu < {\textstyle\frac{1}{2}}, \qquad\mathrm{for} \qquad 0\leq k\lambda_e < 1
\end{equation}
propagating along the Weibel thermal fluctuation field perpendicular to the Weibel anisotropy, i.e. in $\perp$-direction. These are indeed very low frequency Whistlers since $\Omega_e$ based on the thermal fluctuation level is small.

\subsection{Weibel-unstable case}
The interesting domain is that of unstable Weibel modes (for instance in view of application to relativistic shocks) in which case the magnetic field grows and may become quite strong. Several mechanisms of saturation or limiting growth have been proposed \citep[for energy arguments, quasilinear and other nonlinear  mechanisms see][respectively, and references therein]{achterberg2007,pokhotelov2011,pokhotelov2010}. Here we are not interested in saturation but in the excitation of low-frequency plasma modes which may transport the field away from the region of  excitation.

Since $A>0$ for unstable Weibel-like modes growing from thermal background fluctuations, while generating a substantial magnetic field in the direction perpendicular to the direction of anisotropy, we immediately conclude that in the presence of an isotropic background plasma the Whistler mode is naturally unstable under the condition that sufficiently many resonant particles exist with large enough resonant energy. This is certainly the case when Weibel-like modes are excited by a streaming population. In this case one  expects that Whistlers will be radiated along the Weibel-like field perpendicular to the exciting anisotropy direction.

What about the two remaining low-frequency electromagnetic modes, the Firehose and Mirror modes? The former is very easy to infer about. From Eq. (\ref{eq-fire}) and the instability condition $A>0$ we immediately construct the contradiction  $2<0$ thus excluding the Firehose mode from growing in this case. The Firehose mode will not excite any Alfv\'en waves on the expense of the Weibel mode in this unstable case. However, the situation is different for the mirror mode. Clearly Eq. (\ref{eq-mirr}) can be satisfied when the Weibel mode grows with $A>0$ (instead of $A<0$ as in the former Weibel-stable case). The corresponding condition can be written
\begin{equation}
\frac{\beta_\|}{\beta_\perp}-1>\frac{1}{\beta_\|} \qquad\mathrm{and}\qquad A=\frac{\beta_\|}{\beta_\perp}-1>0
\end{equation}
which yields $\beta_\|>1/A$ for mirror instability and can clearly be satisfied. The solution is
\begin{equation}
\beta_\|\gtrsim {\textstyle\frac{1}{2} } \beta_\perp\left(1\pm\sqrt{1+\frac{4}{\beta_\perp^2}} \right)
\end{equation}
Of the two solutions one is trivially satisfied. The other requires that the anisotropy $A>1/\beta_\perp$. It is interesting to note that since the $\beta$s depend inversely on the growing magnetic field, starting from thermal level, they will  decrease with Weibel-like mode growth, thereby ultimately stabilizing the second Mirror mode branch when $A\sim1/\beta_\perp$, while the branch with the negative sign of the square root remains unaffected.

Hence, any growing Weibel-like mode will readily self-consistently evolve into a chain of Mirror structures which on their own are filled with low-frequency Whistlers \citep[for the observational evidence and theoretical arguments see, e.g.,][]{baumjohann1999,treumann2000} bouncing back and force in the mirrors with some of them possibly escaping to the environment.  As a consequence, the Weibel-like instability does not only lead to magnetic and current filaments of perpendicular scale of the order of the electron skin depth. In addition the magnetic Weibel-like fields evolve into a sequence of mirror structures as the lowest frequency magnetically turbulent modes which can evolve according to their own nonlinear dynamics \citep[cf., e.g.,][and references therein]{pokhotelov2008,pokhotelov2010,pokhotelov2013}.

\section{Conclusions}
The Weibel-like instability has frequently been made responsible for the excitation of magnetic fields in otherwise initially non-magnetized plasmas. Actually, as has been known since long time \citep[][and followers]{landau1959,landau1960,sitenko1967, akhiezer1975}, thermal fluctuations in the plasma readily lead to the spontaneous emission of magnetic fluctuations. At the lowest frequencies these form quasi-stationary magnetic fields \citep[see, e.g.,][]{yoon2007,treumann2012} with wavelength longer than the electron skin depth $\lambda_e$. 

These fluctuations provide an initial weak magnetic background field which allows the Weibel-like magnetic field to penetrate over some distance into the otherwise magnetically nontransparent plasma which would inhibit penetration of any magnetic field over distances longer than $\lambda_e$, effectively exponentially screening the plasma from magnetic fields outside the Weibel-like mode source region. In the presence of an pressure anisotropy, however, Firehose, Mirror and Whistler modes can grow in the plasma and compete with the Weibel-like mode. A Weibel-stable plasma allows for Firehose and Whistler modes to grow, both propagating along the magnetic thermal fluctuation field. In a Weibel-unstable plasma the Firehose instability is stable. However, Whistler and Mirror modes can grow. The latter structure the unstably generated Weibel-like magnetic field into chains of magnetic bubbles or holes and trap low frequency Whistlers \citep{baumjohann1999,treumann1999,treumann2000} thus contributing to magnetic structure and turbulence in addition to the known Weibel filamentation effect. This will have a profound effect on the self-consistent generation of magnetic plasma turbulence in regions of pressure anisotropy (or relative streaming) in an otherwise initially non-magnetized plasma. Not only may one expect that anisotropic or streaming plasmas will thus naturally be weakly magnetized, they will also be naturally turbulent thus providing plenty of scattering centers for energetic particles as required in all stochastic particle acceleration scenarios \citep[for a recent review cf., e.g.,][and references therein]{treumann2011,schure2012}. 

The present investigation has been forcefully restricted to the investigation of one quite simple low-frequency electromagnetic mode only, the conventional thermally anisotropic Weibel instability. This instability belongs to a much wider class of low frequency electromagnetic modes which have been investigated in depth only recently \citep{schlick2011a,schlick2011b,felten2013,felten2013a,felten2013b}. Several of these modes generate quasi-stationary magnetic fields in a similar way as the Weibel instability we were dealing with. It will be most interesting to investigate their effect on the excitation of other electromagnetic instabilities, the propagation of Alfv\'en and Whistler waves in the plasma that has become weakly re-magnetized by them. It should also be noted at this occasion that \citet{simoes2013} in a very recent paper performed particle-in-cell simulations of a thermally isotropic, quiescent and stable plasma to determine the electric and magnetic thermal fluctuation levels. As expected, both levels are different from zero. The electric fluctuations naturally map the Langmuir fluctuation branch for wavelength exceeding the Debye length, for shorter wavelengths they show a wide range of weak fluctuations extending even down below the plasma frequency. This result is very well known since long \citep[cf., e.g.,][with and without a beam]{lund1996} and is nicely confirmed by these simulations. The magnetic fluctuations found in this case are also expected in the whole frequency range from fluctuation theory of any thermal system \citep{landau1960,sitenko1967}. Their absence would have been very surprising. As they should, magnetic fluctuation amplitudes maximize at lowest frequencies and longest wavelengths. This was noted by \citet{yoon2007} and, for vanishing anisotropy,  also in \citep{treumann2012}. Both the last two papers became in this respect completely independent of the Weibel mode indicating the general presence of magnetic fluctuations in thermally anisotropic and also thermally isotropic plasmas therby ver softly magnetizing an initially non-magnetic plasma and permitting for propagation of low-frequency electromagnetic (mhd) modes.

\begin{acknowledgements}
This research was part of an occasional Visiting Scientist Programme in 2006/2007 at ISSI, Bern. RT thankfully recognizes the assistance of the ISSI librarians, Andrea Fischer and Irmela Schweizer. RT also thanks the two referees, P. H. Yoon and another anonymous referee, for their intriguing remarks on the paper. One of the referees has brought to our attention the series of papers on the new low frequency relativistic electromagnetic modes which is thankfully acknowledged.
\end{acknowledgements}


\begin{thebibliography}{ }

\bibitem[Achterberg \& Wiersma(2007)]{achterberg2007} Achterberg, A.  \& Wiersma, J.: The Weibel instability in relativistic plasmas. I. Linear theory, II. Nonlinear theory and stabilization mechanism, Astron. Astrophys. 475, 1-18,  doi: 10.1051/0004-6361:20065365, 2007.
 

\bibitem[Akhiezer et al.(1975)]{akhiezer1975} Akhiezer, A. I., Akhiezer, I. A., Sitenko, R. V. \& Stepanov, K. N.: Nonlinear Theory and Fluctuations, in  Plasma Electrodynamics, vol 2, pp. 116-142, Pregamon Press, Oxford, 1975. 

 
\bibitem[Baumjohann \& Treumann(2012)]{treumann1996} Baumjohann, W. \& Treumann, R. A.: Basic Space Plasma Physics, Revised Edition, Imperial College Press at World Scientific, London \& Singapore, 2012. 


\bibitem[Baumjohann et al.(1999)]{baumjohann1999} Baumjohann, W., Treumann, R. A., Georgescu, E., Haerendel, G., Fornacon, K.-H. \& Auster, U.: Waveform and packet structure of lion roars, Ann. Geophys. 17, 1528-1534, doi:10.1007/s00585-999-1528-9, 1999.





\bibitem[Bykov \& Treumann(2011)]{treumann2011} {Bykov, A.~M. \& Treumann, R.~A.: Fundamentals of collisionless shocks for astrophysical application, 2. Relativistic shocks, Astron. Astrophys. Rev. 19, 42, doi: 10.1007/s00159-011-0042-8, arXiv:1105.3221, 2011.} 



















\bibitem[Felten et al.(2013)]{felten2013} Felten T., Schlickeiser R., Yoon P. H. \& Lazar M.: Spontaneous electromagnetic fluctuations in unmagnetized plasmas. II. Relativistic form factors of aperiodic thermal modes, Phys. Plasmas 20, 052113, doi: 10.1063/1.4804402, 2013.

\bibitem[Felten \& Schlickeiser(2013a)]{felten2013a} Felten T. \& Schlickeiser R.: Spontaneous electromagnetic fluctuations in unmagnetized plasmas. IV. Relativistic form factors of aperiodic Lorentzian modes, Phys. Plasmas 20, 082116, doi:10.1063/1.4817804, 2013a.


\bibitem[Felten \& Schlickeiser(2013b)]{felten2013b} Felten T. \& Schlickeiser R.: Spontaneous electromagnetic fluctuations in unmagnetized plasmas. V. Relativistic form factors of weakly damped/amplified thermal modes, Phys. Plasmas 20, 082117, doi:10.1063/1.4817805, 2013b.


\bibitem[Felten \& Schlickeiser(2013c)]{felten2013c} Felten T. \& Schlickeiser R.: Spontaneous electromagnetic fluctuations in unmagnetized plasmas. VI. Transverse, collective mode for arbitrary distribution functions, Phys. Plasmas 20, 104502, doi:10.1063/1.4824114, 2013c.




\bibitem[{{Fried}(1959)}]{fried1959}{Fried} B. D.: {Mechanism for instability of transverse plasma waves}, Phys. Fluids 2, 337-, doi: {10.1063/1.1705933}, 1959.




\bibitem[Landau \& Lifschitz(1959)]{landau1959}Landau L. D. \& Lifshitz E. M., Fluid Mechanics, Pergamon Press, Course of Theoretical Physics, vol. 6, New York, 1959.

\bibitem[Landau \& Lifschitz(1960)]{landau1960}Landau L. D. \& Lifshitz E. M., Electrodynamics of Continuous Media, Pergamon Press, Oxford, 1960.




\bibitem[Lund et al.(1996)]{lund1996} Lund, E.~J., Treumann, R.~A. \& Labelle, J.: Quasi-thermal fluctuations in a beam-plasma system, Phys. Plasmas 3, 1234-1240, 1996, doi:10.1063/1.871747 

















\bibitem[Pokhotelov \& Amariutei(2011)]{pokhotelov2011} Pokhotelov, O. A. \& Amariutei, O. A.: Quasi-linear dynamics of Weibel instability, Ann. Geophys. 29, 1997-2001, doi:10.5194/angeo-29-1997-2011, 2011.

\bibitem[Pokhotelov et al.(2010)]{pokhotelov2010} Pokhotelov, O. A., Sagdeev R.Z., Balikhin M. A., Fedun V. N. \& Dudnikova G. I.: Nonlinear mirror and Weibel modes: peculiarities of quasi-linear dynamics, Ann. Geophys. 28, 2161-2167, doi:10.5194/angeo-28-2161-2010, 2010.
 
\bibitem[Pokhotelov et al.(2013)]{pokhotelov2013} Pokhotelov, O. A., Onishchenko.  O. G. \& Stenflo, L.: Physical mechanisms for electron mirror and field swelling modes, Phys. Scripta 87, ID 065303, doi:10.1088/0031-8949/87/06/065303, 2013.
 
\bibitem[Pokhotelov et al.(2008)]{pokhotelov2008} Pokhotelov, O. A., Sagdeev, R. Z., Balikhin, M. A., Onishchenko.  O. G. \& Fedun, V. N.: Nonlinear mirror waves in non-Maxwellian space plasma, J. Geophys. Res. 113, ID A04225, doi:10.1029/2007JA012642, 2008. 




\bibitem[Schlickeiser et al.(2011)]{schlick2011a} Schlickeiser R., Lazar M. \& Skoda T.: Spontaneously growing, weakly propagating, transverse fluctuations in anisotropic magnetized thermal plasmas., Phys. Plasmas 18, 012103, doi: 10.1063/1.3532787, 2011.

\bibitem[Schlickeiser \& Skoda(2011)]{schlick2011b} Schlickeiser R. \& Skoda T.: Linear theory of weakly amplified, parallel propagating, transverse temperature-anisotropy instabilities in magnetized thermal plasmas, Astrophys. J. 716, 1596-1606, doi: 10.108870004-637X/716/2/1596, 2011.

\bibitem[Schure et al.(2012)]{schure2012} Schure, K. M., Bell, A. R., O'C Drury, L. \& Bykov, A. M.: Diffusive shock acceleration and magnetic field amplification, Space Sci. Rev. 173, 491-519, doi: 10.1007/s11214-012-9871-7, 2012.

\bibitem[Sim$\tilde\mathrm{o}$es et al.(2013)]{simoes2013} Sim$\tilde\mathrm{o}$es, F. J. R., Jr., Pavan, J., Gaelzer, R., Ziebell, L. F. \& Yoon, P. H.: Particle-in-cell simulations on spontaneous thermal magnetic field fluctuations, Phys. Plasmas 20, 100702, doi: 10.1063/1.4825249, 2013.


\bibitem[Sitenko(1967)]{sitenko1967} Sitenko,  A. G.: Electromagnetic Fluctuations in Plasma, Academic Press, New York, 1967.
 










\bibitem[Treumann(2009)]{treumann2009} Treumann, R. A.: Fundamentals of collisionless shocks for astrophysical application, 1. Non-relativistic shocks, Astron. Astrophys. Rev. 17, 409-535, doi: 10.1007/s00159-009-0024-2, 2009.





\bibitem[Treumann \& Baumjohann(2000)]{treumann1999}
Treumann, R. A. \& Baumjohann, W.: Collisionless mirror mode trapping, Nonlin. Process. Geophys. 7, 179-184, 2000.

\bibitem[Treumann \& Baumjohann(1997)]{treumann1997}
Treumann, R. A. and Baumjohann, W.: Advanced Space Plasma Physics, Imperial College Press, London, 1997.

\bibitem[Treumann and Baumjohann(2012)]{treumann2012} Treumann, R. A. and Baumjohann, W.: A note on the Weibel instability and thermal fluctuations, Ann. Geophys. 30, 427-431, doi:10.5194/angeo-30-427-2012, 2012.


\bibitem[Treumann et al.(2000)]{treumann2000}
Treumann, R. A., Georgescu, E. \& Baumjohann, W.: Lion roar trapping in mirror modes, Geophys. Res. Lett. 27, 1843-1846, doi:10.1029/2000GL003767, 2000.



\bibitem[{{Vedenov et al.}(1961)}]{vedenov1961}{Vedenov} A. A., Velikhov E. P. \& Sagdeev, R. Z. :  {Stability of plasma}. Sov. Phys. Uspekhi.  4, 332--369, doi: {10.1070/PU1961v004n02ABEH003341}, 1961.


\bibitem[{{Weibel}(1959)}]{weibel1959}{Weibel} E. S.:  {Spontaneously growing transverse waves in a plasma due to an anisotropic velocity distribution}. Phys. Rev. Lett. 2, 83--84, doi: {10.1103/PhysRevLett.2.83}, 1959.

\bibitem[{{Yoon}(2007)}]{yoon2007}{Yoon} P. H.:  {Spontaneous thermal magnetic field fluctuations}. Phys. Plasmas 14, 83--84,  doi:10.1063/1.2741388, 2007.

\bibitem[{{Yoon \& Davidson}(1987)}]{yoon1987}{Yoon} P. H. \& Davidson R. C.:  {Exact analytical model of the
classical Weibel instability in a relativistic anisotropic plasma}. Phys. Rev. A 35, 2718--2721,  doi:10.1103/PhysRevA.35.2718, 1987.







\end{thebibliography}
\end{document}